\documentclass[
10pt,
aps,
notitlepage, 
bibnotes,
prb,
twocolumn,
showpacs,preprintnumbers,superscriptaddress,
amsmath,amssymb,floatfix]{revtex4-2}

\usepackage{graphicx}
\usepackage{dsfont}
\usepackage{dcolumn}
\usepackage{bm}
\usepackage{color}
\usepackage{xcolor}
\usepackage{url}
\usepackage{tabularx}
\usepackage{array}
\usepackage{booktabs}
\usepackage{comment}
\usepackage{hyperref}
\usepackage{footnote}
\usepackage{lipsum}
\usepackage[normalem]{ulem}
\usepackage{booktabs}

\usepackage{csquotes}
\usepackage{listings}
\definecolor{keyword}{rgb}{0.6,0.2,0.8}  
\definecolor{string}{rgb}{0.2,0.6,0.2}   
\definecolor{comment}{rgb}{0.5,0.5,0.5}  
\definecolor{identifier}{rgb}{0,0,0}     
\definecolor{type}{rgb}{0.2,0.4,0.8}     
\definecolor{operator}{rgb}{0.8,0.1,0.1} 

\lstdefinelanguage{Golang}{
  keywords={break, case, chan, const, continue, default, defer, else, fallthrough, 
            for, func, go, goto, if, import, interface, map, package, range, return, 
            select, struct, switch, type, var},
  keywordstyle=\color{keyword}\bfseries,
  ndkeywords={bool, byte, complex64, complex128, error, float32, float64, int, 
              int8, int16, int32, int64, rune, string, uint, uint8, uint16, 
              uint32, uint64, uintptr, true, false, iota, nil},
  ndkeywordstyle=\color{type}\bfseries,
  identifierstyle=\color{identifier},
  sensitive=true,
  comment=[l]{//},
  morecomment=[s]{/*}{*/},
  commentstyle=\color{comment}\itshape,
  stringstyle=\color{string}\ttfamily,
  morestring=[b]',
  morestring=[b]"
}

\lstdefinestyle{customc}{
  belowcaptionskip=1\baselineskip,
  breaklines=false,
  frame=L,
  xleftmargin=\parindent,
  language=C,
  showstringspaces=false,
  basicstyle=\footnotesize\ttfamily,
  keywordstyle=\bfseries\color{green!40!black},
  commentstyle=\itshape\color{purple!40!black},
  identifierstyle=\color{blue},
  stringstyle=\color{red},
}

\definecolor{colorA}{rgb}{0, 0, 1}
\definecolor{colorB}{rgb}{0.5, 0, 0.9}
\definecolor{colorC}{rgb}{0.4, 0, 0.4}
\definecolor{color_green}{rgb}{0, 0.39, 0}

\makeatletter
  \def\my@tag@font{\normalsize}
  \def\maketag@@@#1{\hbox{\m@th\normalfont\my@tag@font#1}}
  \let\amsmath@eqref\eqref
  \renewcommand\eqref[1]{{\let\my@tag@font\relax\amsmath@eqref{#1}}}
\makeatother

\begin{document}

\def\afflux{Department of Physics and Materials Science, University of Luxembourg, L-1511 Luxembourg, Luxembourg}

\title{Stability and Nucleation of Dipole Strings in Uniaxial Chiral Magnets}

\author{Vladyslav~M.~Kuchkin}
\email{vladyslav.kuchkin@uni.lu}
\affiliation{\afflux}

\author{Nikolai~S.~Kiselev}
 \affiliation{Peter Gr\"unberg Institute, Forschungszentrum J\"ulich and JARA, 52425 J\"ulich, Germany}
 
\author{Andreas~Haller}
\affiliation{\afflux}

\author{\v{S}tefan~Li\v{s}\v{c}\'{a}k}
\affiliation{\afflux}

\author{Andreas~Michels}
\affiliation{\afflux}

\author{Thomas~L.~Schmidt}
\affiliation{\afflux}

\date{\today}

\begin{abstract}
We report on the stability of the magnetic dipole string (DS), a three-dimensional magnetic texture formed by two coupled Bloch points with opposite topological charges, separated by an equilibrium distance.
Previous studies demonstrated the stability of such configurations through geometric confinement or coupling with local perturbations in the magnetization field, such as skyrmion strings or dislocations in helical modulations.
Here, we show that, in uniaxial chiral magnets, an isolated DS remains stable in an unperturbed vacuum, thus representing a true three-dimensional soliton. 
The phase diagram illustrates the stability of the DS embedded in the conical or helical phases across a broad range of material parameters and external magnetic fields.
Using the geodesic nudged elastic band method applied to a regularized micromagnetic model, we demonstrate that isolated DSs are protected from collapse by an energy barrier.
Stochastic spin-lattice simulations demonstrate that DSs can spontaneously nucleate during in-field annealing.
This work aims to stimulate the experimental observation of DSs and further exploration of uniaxial chiral magnets. 
\end{abstract}

\maketitle

Magnetic skyrmions~\cite{Bogdanov_89} are two-dimensional (2D) topological solitons that can be stabilized in various magnetic systems~\cite{Blugel2011,vdWaals,Manfred}.
Their unique particle-like properties have garnered significant interest for potential applications~\cite{Tokura2020, Song_20, Panagopoulos_21}.

In bulk 3D samples, magnetic skyrmions can form intricate filamentary textures of nontrivial topology.
For example, 3D skyrmion strings can braid~\cite{braid}, hybridize~\cite{Kuchkin_2022}, cluster into bundles~\cite{bundle}, or even host hopfion rings~\cite{Zheng_23}.
Typically, these strings penetrate the entire thickness of a sample. 
However, there are exceptions, such as configurations in which a skyrmion string emerges from the surface and terminates within the sample volume by a singularity -- a so-called Bloch point (BP)~\cite{Feldtkeller_1965,Doring_1968}.
These statically stable configurations are referred to as chiral bobbers~\cite{Rybakov_2015, Zheng_18, Lux}.  

\begin{figure}[tb!]
\centering
\includegraphics[width=7.8cm]{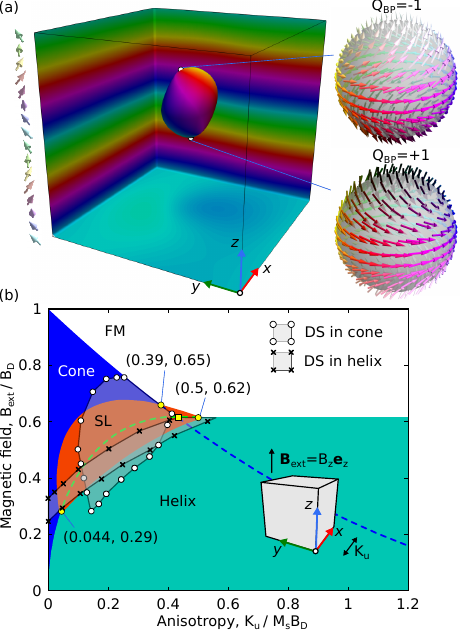}
\caption{~\small (a)~Stable DS solution in the cone phase obtained by direct energy minimization of Eq.~\eqref{mag_energy} at $h= B_{\rm ext}/B_{\rm D} = 0.55$ and $u =\mathcal K_{\rm u}/(M_{\rm s} B_{\rm D})= 0.25$, visualized via the isosurface ($n_z = 0$). 
The inset highlights two Bloch points (BPs) with opposite topological charges~\cite{Malozemoff_79}.
The standard color code~\cite{Magnoom, Savchenko} is used throughout: white and black pixels indicate magnetic moments aligned parallel and antiparallel to the $z$~axis, respectively, while red, green, and blue represent the azimuthal angle relative to the $x$~axis.  
(b) State diagram for chiral magnets under an external field applied orthogonally to the hard axis of the magnetocrystalline anisotropy.
The red region corresponds to the skyrmion lattice (SL), and the dashed green line marks the conditions where the metastable cone and helix phases have identical energies.
The dashed blue line, instead, marks the critical field at which the cone saturates to a FM state.
The two semitransparent regions indicate the stability domains of DSs within the cone and helix phases.
} 
\label{Fig1}
\end{figure}

The magnetic textures studied in this work can be viewed as fragments of skyrmion strings that begin and terminate within the crystal volume, encompassing two BPs, as illustrated in Fig.~\ref{Fig1}(a). 
In the literature, these configurations are referred to as magnetic dipole strings (DSs)~\cite{quasimonopoles}, torons~\cite{toron}, cocoons~\cite{cocoon}, globules~\cite{globule}, or monopole-antimonopole pairs~\cite{Achim, Jiadong}.  

Previous studies have shown that the stabilization of DSs can be achieved through geometric confinement~\cite{cylinder}, artificial surface anisotropy~\cite{Leonov_2018,Leonov_2021}, or coupling to auxiliary textures, such as skyrmion strings, conical edge dislocations~\cite{quasimonopoles}, or conical screw dislocations~\cite{Azhar}.
However, DS configurations stabilized by these methods are constrained by their environments and do not represent true 3D magnetic solitons capable of free motion in all spatial directions.
In this Letter, we introduce a class of magnetic crystals where DSs emerge as freely mobile, true 3D solitons stabilized purely by local interactions. 

\begin{figure*}[tb!]
\centering
\includegraphics[width=17.9cm]{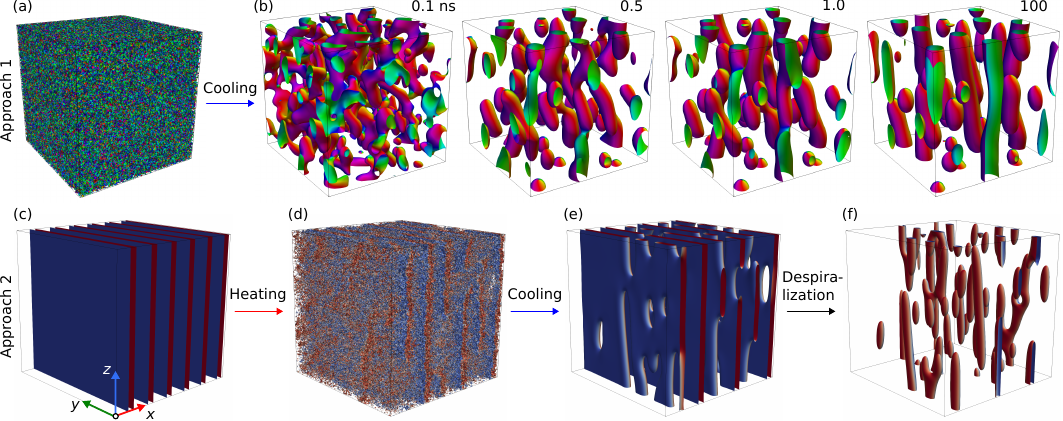}
\caption{~\small Magnetic dipole string (DS) nucleation via in-field annealing (a$–$b) and helix breaking (c$–$f).  
(a)~Random spin distribution resembling a paramagnetic state.  
(b)~Sequence of states obtained using LLG simulations at $h = 0.55$ and $u = 0.25$, starting from~(a). 
The simulated box size is $3L_\mathrm{D} \times 3L_\mathrm{D} \times 3L_\mathrm{D}$, discretized into $128^3$ cuboids.  
(c)~Helix initial state at $h = 0.57$ and $u = 0.42$. 
(d)~Helix at a temperature of $T \sim 0.9T_\mathrm{C}$. 
(e)~State obtained after cooling from (d), showing DSs embedded within the helicoidal phase. 
(f)~Same as~(e) but after the despiralization procedure, which unwinds the helix into a ferromagnetic state.  
The simulated domain in (c$–$f) is $6L_\mathrm{D} \times 6L_\mathrm{D} \times 6L_\mathrm{D}$, discretized into $256^3$ lattice sites. 
The red-blue color scale in (c$–$f) represents the $n_y$~component of the magnetization. Periodic boundary conditions are applied in all directions.
} 
\label{Fig2}
\end{figure*}

In particular, we consider a chiral ferromagnet with bulk-type Dzyaloshinskii-Moriya interaction~\cite{Dzyaloshinskii,Moriya} (DMI) and hard-axis (easy-plane) anisotropy:
\begin{equation}
    \mathcal{E}(\mathbf{n})=\intop\left[\mathcal{A}(\partial_{i} \mathbf{n})^{2}+\mathcal{D}\,\mathbf{n}\cdot\nabla\times\mathbf{n}+U(\mathbf{n})\right]\mathrm{d}V,\label{mag_energy}
\end{equation}
where $\mathcal{A}$ and $\mathcal{D}$ are the strengths of exchange and DMI, respectively; summation over subscript $i\in\{ x, y, z \}$ is assumed.
The potential term $U(\mathbf{n})= - \frac{1}{2} M_\mathrm{s}\mathbf{B}_\mathrm{d}\cdot\mathbf{n}-M_\mathrm{s}B_\mathrm{ext} n_z+\mathcal{K}_\mathrm{u}n_x^{2}$ accounts for the demagnetizing field $\mathbf{B}_\mathrm{d}$, the externally applied magnetic field $B_\mathrm{ext}>0$ along the $z$~axis, and an easy-plane anisotropy $\mathcal{K}_\mathrm{u}>0$ with the hard axis along the $x$~axis.

For further convenience, we introduce the characteristic length scale $L_\mathrm{D}=4\pi\mathcal{A}/\mathcal{D}$ and the magnetic field $B_\mathrm{D} = \mathcal{D}^{2}/(2\mathcal{A}M_\mathrm{s})$ that correspond to the equilibrium period of the chiral modulations at $U(\mathbf{n})=0$ and to the saturation magnetic field at $\mathbf{B}_\mathrm{d}=0$ and $\mathcal{K}_\mathrm{u}=0$, respectively.
Using these characteristic parameters, we can employ the dimensionless field $h=B_\mathrm{ext}/B_\mathrm{D}$ and anisotropy $u=\mathcal{K}_\mathrm{u}/(M_\mathrm{s}B_\mathrm{D})$.
In the following discussion, we ignore the presence of demagnetizing fields that usually introduce quantitative but not qualitative changes to most of the magnetic spin textures studied previously~\cite{braid,Rybakov_2015,Kuchkin2023}, especially for the bulk.

Figure~\ref{Fig1}(b) presents the diagram of magnetic states for the model \eqref{mag_energy} in a scenario~\cite{Karhu}, where the external field is applied perpendicular to the hard axis (see bottom right inset).
The ground state of the system is the spin spiral. 
It is common to distinguish spin spirals based on the orientation of their wave vector $\mathbf{q}$ relative to the applied magnetic field. In the case of $\mathbf{q} \parallel \mathbf{B}_\mathrm{ext}$, the spin spiral is referred to as the \textit{cone}, while for $\mathbf{q} \perp \mathbf{B}_\mathrm{ext}$, it is typically referred to as the \textit{helix}.
The cone and helix can serve as a vacuum for ordinary and hybrid skyrmion strings~\cite{braid,Kuchkin_2022}, heliknotons~\cite{Voinescu, Kuchkin2023}, and 3D chiral droplets~\cite{Kuchkin_2022}.
In Fig.~\ref{Fig1}(b), the regions of stable DSs submerged in the cone or helix phases are depicted by the gray semi-transparent areas marked by circles and triangles, respectively.

Let us briefly discuss the key aspects of the diagram shown in Fig.~\ref{Fig1}(b).
In the isotropic case ($u=0$), there are only two phases: the cone phase for $h<1$ and the saturated or ferromagnetic (FM) state for $h>1$.
Similarly to the case of an easy-axis anisotropy~\cite{Leonov2017}, for hard-axis bulk chiral magnets, the phase diagram contains four phases~\cite{Karhu}: cone, helix, skyrmion lattice (SL), and FM [Fig.~\ref{Fig1}(b)]. 
The equilibrium period of the cone phase in the case $u>0$ depends on the magnetic field $h$ and cannot be found analytically.
On the other hand, by analyzing the Euler-Lagrange equations, one can derive the critical field $h_{1}^\mathrm{cr}$ at which the system saturates to the FM state:
\begin{equation}
h_{1}^\mathrm{cr}=\left(1-u/2\right)^{2},\,\,u\in[0,2].\label{critical_field}
\end{equation}
For details, we refer the reader to the Supplementary Material~\ref{AppendixA}. In the diagram Fig.~\ref{Fig1}(b), the function $h_{1}^\mathrm{cr}$$(u)$ is marked by the dashed blue line.

The period of the helix is a function of $h$, but in our geometry, when $\mathbf{B}_\mathrm{ext}$ is orthogonal to the hard anisotropy axis, it does not depend on $u$.
Thus, the transition to the FM state can be found by analyzing a single $360^{\circ}$-domain wall state that corresponds to $h_{2}^\mathrm{cr}=\pi^{2}/16$.
Comparing the latter with \eqref{critical_field}, we find a critical anisotropy value of $u^\mathrm{cr}=1-\pi/2$, corresponding to the triple point -- cone-helix-FM (see square symbol in the diagram). 

To investigate phase transitions between the cone, helix, and SL phases, we numerically minimized the Hamiltonian~\eqref{mag_energy} using optimized parameters for each phase and compared their respective energies.
These calculations were performed with the conjugate gradient method, enabling high-precision estimates of phase transitions and stability of the DS.
Specifically, we identified three triple points corresponding to the cone-helicoid-SL, cone-FM-SL, and helicoid-FM-SL transitions, which in terms of $(u, h)$ are located at $(0.044, 0.29)$, $(0.39, 0.65)$, and $(0.5, \pi^{2}/16)$, respectively.

For the study of the stability region of the DS, we constructed a reliable ansatz of its profile that can be used as an initial guess in numerical simulations.
Using the parameterization $\mathbf{n}=(\sin\Theta\cos\Phi,\sin\Theta\sin\Phi,\cos\Theta)$, the ansatz for the DS can be written in bispherical coordinates.
These coordinates are characterized by two poles at $(0,0,\pm a/2)$, which coincide with the BP positions:
\begin{equation}
    \!\!\!\!\Theta = \arccos\left[\dfrac{r^{2}\!-\!a^{2}}
    {\sqrt{\left(r^{2}\!+\!a^{2}\right)^{2}\!-\!4a^{2}z^{2}}}\right]\!\!,\,\Phi = \arctan\dfrac{y}{x}\!+\!\dfrac{\pi}{2}, \label{dipole_string_ansatz}
\end{equation}
where $r\!=\!\sqrt{x^{2}\!+\!y^{2}\!+\!z^{2}}$ denotes the spherical radius and $a$ is the distance between the poles (between the BPs).

Since Eq.~\eqref{dipole_string_ansatz} gives $\mathbf{n} = (0, 0, 1)$ at $r \rightarrow \infty$, it is necessary to rotate both the vector field $\mathbf{n}$ and the spatial coordinates $\mathbf{r}$ to obtain the cone phase vacuum, as described in Ref.~\cite{Haifeng_2018}.
In the case where the DS is embedded in the helix, we place the BPs at $(\pm a/2,0,0)$ and apply the spiralization procedure~\cite{Kuchkin2023}, which transforms the FM state into a helix with $\mathbf{q}\parallel \mathbf{e}_x$.
For both cases, i.e., DS in the helix and DS in the cone, we provide Mumax3 scripts in Supplementary Material~\ref{Mumax_cone}, ~\ref{Mumax_helicoid}, where the ansatz \eqref{dipole_string_ansatz} is implemented.

Starting the energy minimization with the above ansatz, one can obtain a statically stable solution for a DS when $B_\mathrm{ext}$ and $\mathcal{K}_\mathrm{u}$ are in the ranges depicted as semi-transparent gray areas in Fig.~\ref{Fig1}(b). 
Varying the values of $B_\mathrm{ext}$ and $\mathcal{K}_\mathrm{u}$, we identify the boundaries of the DS stability regions. 
Note that the DS within the cone and helix phases in Fig.~\ref{Fig1}(b) represent metastable solutions.
This implies that while an isolated DS is a statically stable configuration in these regions, the total energy of the system with a DS is higher than that of the pure cone or helix state.
Interestingly, even in the case of the isotropic chiral magnet ($u=0$), the DS can be stabilized inside the helicoidal phase in a finite range of the applied field, $ h \in [0.26, 0.33]$.
In this scenario, the DSs are stabilized in a helix, while the ground state of the system is the cone phase. 
Nevertheless, this indicates that DSs can be experimentally observed even in isotropic chiral magnets, e.g., FeGe~\cite{Yu_11, Kov17, Du_18, Yu_18}, MnSi~\cite{Muhlbauer_09,Yu_15}, Fe$_{1-x}$Co$_{x}$Si~\cite{Yu_10, Park_14}, and other B20-type crystals~\cite{kanazawa2017}.

We consider two approaches to demonstrate the nucleation ability of DSs in the cone and helix phases (Fig.~\ref{Fig2}).
The first approach illustrates DS nucleation via annealing.
We begin with a random spin distribution that resembles the paramagnetic phase [Fig.~\ref{Fig2}(a)].
The system is then cooled using standard Landau-Lifshitz-Gilbert (LLG) simulations, resulting in the configurations shown in Fig.~\ref{Fig2}(b).  
This simulation was performed with Mumax~\cite{Mumax} using a cubic sample and periodic boundary conditions.  
After long-term relaxation, the system converges to a mixed state consisting of skyrmion strings and several DSs of different sizes.
Similar configurations are consistently observed when the system is cooled from different initial random states.
In these simulations, we frequently observe stable DSs of various lengths corresponding to different equilibrium distances between the BPs.
Such behavior of DSs coupled to skyrmion strings has been reported previously~\cite{quasimonopoles}. 
It is important to note that from a topological perspective, all these DSs are identical to the one depicted in Fig.~\ref{Fig1}(a).

The second approach resembles the method used to nucleate monoaxial skyrmions~\cite{mono_sk} in experiments~\cite{Du_22}.
Starting from a helical state [Fig.~\ref{Fig2}(c)], we perform stochastic LLG simulations using a spin-lattice model and the semi-implicit method~\cite{Mentink_10}.
In these simulations, the temperature is set to $T = 1.2 J/k_\mathrm{B}$, where $J = 2 \mathcal{A}$ represents the Heisenberg exchange constant.
This temperature is below the Curie temperature of $T_\mathrm{C} = 1.345 J/k_\mathrm{B}$, as estimated in Ref.~\cite{quasimonopoles}.
After some time, we observed the breaking of several spirals [Fig.\ref{Fig2}(d)].
The temperature is then turned off, and the system is gradually cooled.
The resulting final state consists of several DSs embedded within the helicoidal state [Fig.~\ref{Fig2}(e)].
To enhance visualization, we applied the despiralization procedure, as described in Ref.~\cite{Kuchkin2023} [Fig.~\ref{Fig2}(e)].
Similarly to the first approach, the final state typically comprises skyrmion strings and multiple DSs of varying lengths.

\begin{figure*}
\centering
\includegraphics[width=18cm]{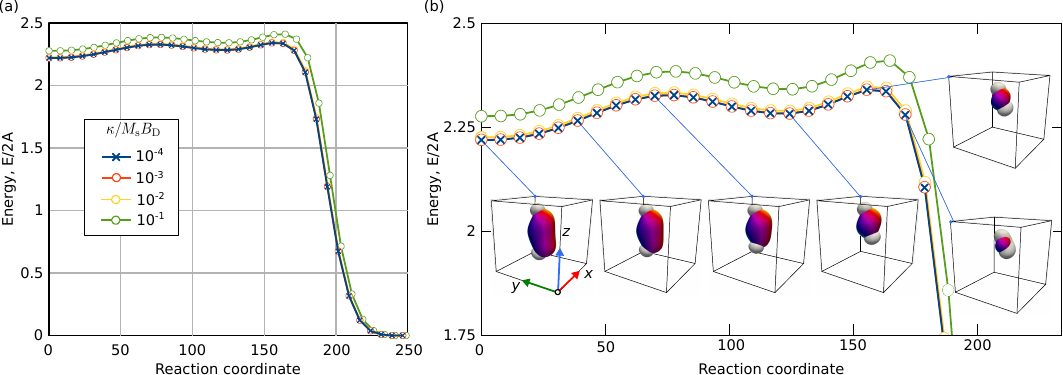}
\caption{~\small (a)~Minimum energy path (MEP) for dipole string (DS) collapse, shown for different values of the parameter $\kappa$ in the regularized micromagnetic model. 
The path connects the stable DS state (leftmost point) to the cone phase state (rightmost point) via the DS's shrinking.
(b)~Zoomed-in fragment of the MEP from~(a), with selected DS configurations visualized at various path images. 
Isosurfaces correspond to $n_z = 0$, while the white surfaces represent $|\mathbf{n}| = 0.95$.
The MEP calculations were performed at $h = 0.55$, $u = 0.25$ in a simulated domain of size $3 L_\mathrm{D} \times 3L_\mathrm{D} \times 3L_\mathrm{D}$, discretized with 21 nodes per $ L_\mathrm{D}$. The cubic domain in the insets has a side length of $ 1.5L_\mathrm{D}$. 
} 
\label{Fig3}
\end{figure*}

For the further characterization of DSs, we estimated the energy barriers that protect them from collapse. 
A representative minimum energy path (MEP), calculated using the geodesic nudged elastic band (GNEB) method~\cite{bessarab_2015,bessarab_2017}, is shown in Fig.~\ref{Fig3}. 
We found that within the standard micromagnetic model where magnetization vectors \(\mathbf{n}\) are defined on the \(\mathbb{S}^2\) sphere (i.e., ordinary three-dimensional unit vectors), the GNEB method exhibits poor convergence for configurations containing BPs. 
This issue arises due to the divergence of the effective field in the vicinity of the BP. 
To address this, we used a regularized micromagnetic model, where the order parameter $\boldsymbol{\nu}$ is defined on the $\mathbb{S}^3$ sphere~\cite{Kuchkin2024}.  
The first three components of the four-dimensional vector $\boldsymbol{\nu} = (\nu_1, \nu_2, \nu_3, \nu_4)$ coincide with the magnetization unit vector $\mathbf{n}$, while the fourth component relates to the vector's length:  
\begin{equation}
    \nu_{1}=n_x, \nu_{2}=n_y, \nu_{3}=n_z, \nu_{4}^{2}=1-|\mathbf{n}|^2.\label{nu}
\end{equation}
In this model, the exchange energy term in Eq.~\eqref{mag_energy} is modified as follows:
\begin{equation}
   \mathcal{A}\sum_{i=x,y,z}(\partial_{i} \mathbf{n})^{2} \mapsto \mathcal{A}\sum_{i=x,y,z}(\partial_{i} \bm{\nu})^{2} + \kappa \nu_{4}^{2},\label{gen_ham}
\end{equation}
while the DMI and potential energy term remain unchanged. 
The phenomenological parameter $\kappa$ in Eq.~\eqref{gen_ham} can be interpreted as a BP localization parameter.  
In the limit $\kappa\rightarrow\infty$ ($\nu_4\rightarrow 0$), the system reduces to the standard micromagnetic model.
The regularized model inherently satisfies the condition $|\mathbf{n}| \leq 1$, ensuring qualitative consistency with the predictions of the more general quantum-mechanical model~\cite{Kuchkin2024}.

The key aspects of the regularized GNEB (RGNEB) method remain identical to those of the standard GNEB method.
The entire transition path is discretized into a series of transient states, referred to as images, which interact with each other through forces.
The total force acting on each image consists of the effective magnetic field force $\bm{\beta}_{\mathcal{I}}$ and the spring force $\bm{f}_\mathcal{I}$:
\begin{eqnarray}
    \bm{\beta}_\mathcal{I} = -\dfrac{1}{M_\mathrm{s}}\dfrac{\delta\mathcal{H}}{\delta\bm{\nu}_{\mathcal{I}}}, \,\,\,\,\,\, \bm{f}_\mathcal{I} = k(\rho_{\mathcal{I}+1,\mathcal{I}}-\rho_{\mathcal{I},\mathcal{I}-1})\bm{\tau}_{\mathcal{I}},
\end{eqnarray}
where $\mathcal{I}\in[1,\mathcal{N}]$ is the image index with $\mathcal{N}$ being the total number of images, $k$ is the spring force constant, $\rho$ is the reaction coordinate -- the Euclidian distance between the images, and $\bm{\tau}_{\mathcal{I}}$ is the tangent vector. 
The optimal minimum energy path (MEP) is obtained through iterative minimization of the absolute value of the  total force.
Details of the RGNEB implementation, including the force minimization algorithm, are provided in the Supplemental Material~\ref{AppendixB}, ~\ref{AppendixC}.

Figure~\ref{Fig3}(a) shows the MEP calculated for four different values of $\kappa$.  
As $\kappa$ increases and the model approaches the classical micromagnetic model, the energy at each point along the MEP increases while the convergence of the RGNEB method slows down significantly.
For $\kappa > 0.1$, achieving reasonable accuracy becomes problematic, as the RGNEB method fails to converge due to the pinning of the BPs in the lattice.

The initial path guess was generated using the ansatz~\eqref{dipole_string_ansatz}, where the distance between the BPs, $a$, changes linearly along the MEP. 
For all values of $\kappa$, the MEP exhibits two saddle points and one intermediate minimum, corresponding to a more compact but slightly higher energy DS.  
A zoomed-in fragment of the MEP is shown in Fig.~\ref{Fig3}(b), along with system snapshots for selected images where the DS is visualized using two isosurfaces.
The color-coded isosurface indicates the shape and size of the DS; the second one highlights the BP positions.
The full set of snapshots is available in Supplementary Movie~1.

To conclude, in this work, we introduced a model of a bulk chiral magnet where a magnetic dipole string (DS) exists as a truly statically stable soliton.
We demonstrated that DSs can be stabilized within the cone or helix phases, or in mixed states coexisting with skyrmion strings.
It is shown that uniaxial chiral magnets are promising systems for the experimental observation of DSs when the external magnetic field is applied perpendicular to the hard axis. However, we also show that DSs can be stabilized in isotropic chiral magnets in a certain range of applied magnetic fields.
Furthermore, we argue that spontaneous nucleation of DSs can be achieved through controlled annealing of the sample, making this phenomenon experimentally accessible.
Finally, we demonstrate that a regularized micromagnetic model provides an adequate framework for describing DSs and can be readily adapted for minimum energy path calculation methods, facilitating further theoretical and computational studies.

\begin{acknowledgments}
The authors acknowledge financial support from the National Research Fund of Luxembourg under Grant No.~C22/MS/17415246/DeQuSky. 
N.S.K.\ acknowledges support from the European Research Council under the European Union's Horizon 2020 Research and Innovation Programme (Grant No.~856538---project ``3D MAGiC'').
\end{acknowledgments}

\newpage

\onecolumngrid
\newpage
\begin{center}
   \textbf{Supplemental Material for ``Stability and Nucleation of Dipole Strings in Uniaxial Chiral Magnets''}\newline 
\end{center}

\section{Cone-FM transition}\label{AppendixA}
By performing the nondimensionalization in Hamiltonian (1) as described in the main text, one can obtain the functional in the following form:
\begin{eqnarray}
    \dfrac{\mathcal{E}(\mathbf{n})}{2\mathcal{A}}=\intop\left[\dfrac{(\nabla \mathbf{n})^{2}}{2}+2\pi\mathbf{n}\cdot\nabla\times\mathbf{n}+4\pi^{2}\left(u n_x^{2}-h n_z\right)\right]\mathrm{d}V.
    \label{mag_energy_e0}
\end{eqnarray}
To satisfy the constraint $|\mathbf{n}|=1$, we can employ parametrization with the spherical angles $(\Theta,\Phi)$ as $\mathbf{n}=(\sin\Theta\cos\Phi,\sin\Theta\sin\Phi,\cos\Theta)$. In the case of the cone, the angles depend only on the $z$~coordinate.
The Euler-Lagrange (EL) equations followed from $\delta\mathcal{E}=0$ have the form:
\begin{equation}
    \begin{cases}
        \left(\frac{1}{2}\left(\Phi^{\prime}\right)^{2}-2\pi\Phi^{\prime}-4\pi^{2}u\cos^{2}\Phi\right)\sin2\Theta-\Theta^{\prime\prime}+4\pi^{2}h\sin\Theta=0,\\
        \left(4\pi^{2}u \sin2\Phi-\Phi^{\prime\prime}\right)\sin^{2}\Theta-\Theta^{\prime}\left(\Phi^{\prime}-2\pi\right)\sin2\Theta=0.
    \end{cases}\label{EL_TF}
\end{equation}
For $u=0, h\in[0,1]$, the analytic solution for the cone phase writes as $\mathbf{n}_\mathrm{c}=(\sin\Theta_\mathrm{c}\cos(2\pi z+\phi_0),\sin\Theta_\mathrm{c}\sin(2\pi z+\phi_0),\cos\Theta_\mathrm{c})$, where $\cos\Theta_\mathrm{c}=h$ and $\phi_0$ may take arbitrary value.
At $u>0$, the analytic solution of \eqref{EL_TF} corresponding to the cone phase is not known and can be found only numerically. 
However, the critical magnetic field corresponding to the transition to the FM state can be found analytically. 
To obtain this solution, we parametrize magnetization with stereographic projections, $(\gamma_{1},\gamma_2)$, as:
\begin{eqnarray}
    \mathbf{n} = \left(\dfrac{2\gamma_{1}}{1+\gamma_{1}^{2}+\gamma_{2}^{2}},\dfrac{2\gamma_{2}}{1+\gamma_{1}^{2}+\gamma_{2}^{2}},\dfrac{1-\gamma_{1}^{2}-\gamma_{2}^{2}}{1+\gamma_{1}^{2}+\gamma_{2}^{2}}\right).\label{stereog_proj} 
\end{eqnarray}
In these projections, the solution for the cone phase at $u=0$ can be written as:
\begin{equation}
    \gamma_{1}+i\gamma_{2}=e^{i\left(2\pi z+\phi_{0}\right)}\tan\dfrac{\Theta_\mathrm{c}}{2}.\label{cone_phase_gamma}
\end{equation}
From this, we can deduce, at the transition to the FM state, $h=h^{\rm cr}$, one has $\Theta_\mathrm{c}\rightarrow0$, or equivalently, $\gamma_{1}\rightarrow0$ and $\gamma_{2}\rightarrow0$. 
Plugging \eqref{stereog_proj} into \eqref{mag_energy_e0}, we derive the EL equations for $\left(\gamma_{1},\gamma_{2}\right)$ and linearize them at $\gamma_{1}\ll1$ and $\gamma_{2}\ll1$:
\begin{eqnarray}
    &&4\pi^{2}\left(h^\mathrm{cr}-2u\right)\gamma_{1}-4\pi\gamma_{2}^{\prime}-\gamma_{1}^{\prime\prime}=0,\nonumber\\
    &&4\pi^{2}h^\mathrm{cr}\gamma_{2}+4\pi\gamma_{1}^{\prime}-\gamma_{2}^{\prime\prime}=0,\label{eq_g1g2}
\end{eqnarray}
which is equivalent to a single higher-order linear equation:
\begin{equation}
\!\!\!\gamma_{1}^{(4)}+8\pi^{2}\left[2-h^\mathrm{cr}+u\right]\gamma_{1}^{\prime\prime}+16\pi^{4}h^\mathrm{cr}\left(h^\mathrm{cr}-2u\right)\gamma_{1}=0.\label{eq_g1}
\end{equation}
Characteristic polynomial for \eqref{eq_g1}:
\begin{equation}
    \lambda^{4}+8\pi^{2}\left[2-h^\mathrm{cr}+u\right]\lambda^{2}+16\pi^{4}h^\mathrm{cr}\left(h^\mathrm{cr}-2u\right)=0,\label{polynom}
\end{equation}
has four solutions:
\begin{equation}
\lambda=\pm2\pi\sqrt{h^\mathrm{cr}-2-u\pm\sqrt{\left(2+u\right)^{2}-4h^\mathrm{cr}}},\label{lambda}
\end{equation}
where signs $\pm$ in front of both square roots have to be treated separately.
The instability criterion for the cone phase is obtained by setting the expression under the inner square root to zero and reads:
\begin{equation}
    h^\mathrm{cr}=\left(1-\dfrac{u}{2}\right)^{2}, \, u\in[0,2].
\end{equation}
The latter equation means one can find a stable cone phase for all $u\in[0,2]$ and $0\leq h<h^\mathrm{cr}$.

\section{Mumax3 script for DS ansatz in cone phase}\label{Mumax_cone}

\begin{lstlisting}
OutputFormat = OVF2_BINARY
/************ Material constants (FeGe) ************************/
Ms := 384e3; Msat = Ms 	// saturation magnetization [A/m]
LD := 70.0e-9 			    // spiral period            [nm]
A := 4.0e-12 			      // exchange stifness        [J/m]
DMI := 4.0*pi*A/LD 		  // DMI                      [J/m^2]
Aex = A
Dbulk = DMI
BD := DMI * DMI / (2 * A * Ms)
/******* Easy-plane anisotropy and External magnetic field *****/
anisU = vector(1, 0, 0)
Ku1 = -0.25 * 2 * DMI * DMI / (4 * A)
B := 0.55 * BD
B_ext = vector(0, 0, B)
// Ku1 and B_ext correspond to equilibrium spiral period: 
// L = 1.016232685 * LD;
/********** Sets GridSize, CellSize and PBC ********************/
n := 128;
SetGridSize(n, n, n);
L := 3.0 * 1.016232685 * LD;
d := L/n; 
SetCellSize(d, d, d);
EnableDemag = false
SetPBC(1, 1, 1)
openbc = false
// /**************** Initial state *****************************/
// Globule ansatz in bispherical coordinates
aBP := 0.3; //distance between Bloch points
Th  := 0.8;	 //cone phase angle
for ix:=0; ix<n; ix++{
    for iy:=0; iy<n; iy++{
        for iz:=0; iz<n; iz++{
            rx := -1.5 + 3.0*ix/n + 1e-6 //to exclude division by 0
            ry := -1.5 + 3.0*iy/n + 1e-6 //to exclude division by 0
            rz := -1.5 + 3.0*iz/n + 1e-6 //to exclude division by 0
            r := sqrt(rx*rx+ry*ry+rz*rz);
            f := atan2(ry,rx)
            Q := sqrt((r*r+aBP*aBP)*(r*r+aBP*aBP) - 4.0*rz*rz*aBP*aBP);
            theta := acos((r*r-aBP*aBP)/Q);
            phi   := f + 0.5*pi - 2.0*pi*rz;
            mx := sin(theta)*cos(phi)*cos(Th) - cos(theta)*sin(Th);
            mz := cos(theta)*cos(Th) + sin(theta)*cos(phi)*sin(Th);
            my := sin(theta)*sin(phi);
            m1 := mx*cos(2.0*pi*rz) - my*sin(2.0*pi*rz);
            m2 := my*cos(2.0*pi*rz) + mx*sin(2.0*pi*rz);
            m3 := mz;
            mnew:=vector(m1, m2, m3);
            m.SetCell(ix, iy, iz, mnew);
        }                                  
    }
}
/**************** Running minimization ***********************/
save(m)
relax()
minimize()
save(m)
\end{lstlisting}

\newpage 

\section{Mumax3 script for DS ansatz in helicoidal phase}\label{Mumax_helicoid}

\begin{lstlisting}
OutputFormat = OVF2_BINARY
/************ Material constants (FeGe) ************************/
Ms := 384e3; Msat = Ms 	// saturation magnetization [A/m]
LD := 70.0e-9 			    // spiral period            [nm]
A := 4.0e-12 			      // exchange stiffness       [J/m]
DMI := 4.0*pi*A/LD 		  // DMI                      [J/m^2]
Aex = A
Dbulk = DMI
BD := DMI * DMI / (2 * A * Ms)
/******* Easy-plane anisotropy and External magnetic field *****/
anisU = vector(1, 0, 0)
Ku1 = -0.42 * 2 * DMI * DMI / (4 * A)
B := 0.57 * BD
B_ext = vector(0, 0, B)
// Ku1 and B_ext correspond to equilibrium spiral period: 
// L = 1.371152 * LD;
/********** Sets GridSize, CellSize and PBC ********************/
n := 128;
SetGridSize(n, n, n);
L := 3.0 * 1.371152 * LD;
d := L/n; 
SetCellSize(d, d, d);
EnableDemag = false
SetPBC(1, 1, 1)
openbc = false
// /**************** Initial state *****************************/
// Globule ansatz in bispherical coordinates
aBP := 0.4; //distance between Bloch points
for ix:=0; ix<n; ix++{
    for iy:=0; iy<n; iy++{
        for iz:=0; iz<n; iz++{
            rx := -1.5 + 3.0*ix/n + 1e-6 //to exclude division by 0
            ry := -1.5 + 3.0*iy/n + 1e-6 //to exclude division by 0
            rz := -1.5 + 3.0*iz/n + 1e-6 //to exclude division by 0
            r := sqrt(rx*rx+ry*ry+rz*rz);
            f := -atan2(ry,rz)
            Q := sqrt((r*r+aBP*aBP)*(r*r+aBP*aBP) - 4.0*rx*rx*aBP*aBP);
            theta := acos((r*r-aBP*aBP)/Q);
            phi   := f + 0.5*pi;
            mx := sin(theta)*cos(phi);
            my := sin(theta)*sin(phi);
            mz := cos(theta);
            m1 := mx
            m2 := -my*cos(2*pi*rx)+mz*sin(2*pi*rx)
            m3 := -mz*cos(2*pi*rx)-my*sin(2*pi*rx)
            mnew:=vector(m1, m2, m3);
            m.SetCell(ix, iy, iz, mnew);
        }                                  
    }
}
/**************** Running minimization ***********************/
save(m)
relax()
minimize()
save(m)
\end{lstlisting}

\newpage
\section{Details of the regularized GNEB method}\label{AppendixB}
Let us denote the set of images (transient states) along the path as  $\mathcal{M} = \{\mathcal{M}_\mathcal{I}, \mathcal{I}\in[1,\mathcal{N}]\}$, where $ \mathcal{M}_\mathcal{I} = \{\bm{\nu}_{i,\mathcal{I}} \,|\, i \in [1, N]\}$ represents the \(\mathcal{I}\)-th of \(\mathcal{N}\) images. 
Each image contains $N$ four-dimensional vectors $ \bm{\nu} $, which define the state.
We define the distance between two images, $\mathcal{I}$ and $\mathcal{J}$, as the Euclidean distance:
\begin{equation}
    \rho_{\mathcal{I},\mathcal{J}}=\sqrt{\displaystyle{\sum_{i=1}^{N}\sum_{j=1}^{4}}\left(\nu_{\mathcal{I},i,j}-\nu_{\mathcal{J},i,j}\right)^{2}}.
\end{equation}
In practice, we need to know the distance $\rho$ between neighboring images only, \textit{i.e.}, $\mathcal{J}=\mathcal{I}-1$ or $\mathcal{J}=\mathcal{I}+1$. 
All intermediate images, $1 < \mathcal{I} < \mathcal{N}$, are \textit{movable} and must be optimized using the algorithm described below. 
Meanwhile, the first $\mathcal{I} = 1$ and last $\mathcal{I} = \mathcal{N}$ images are stationary states corresponding to the local minima of the Hamiltonian.
For further convenience, we define the following projection operator:
\begin{equation}
    \mathcal{P}(\mathbf{a},\mathbf{b})=\mathbf{b}-(\mathbf{a}\cdot\mathbf{b})\mathbf{a},
\end{equation}
which subtracts from a vector $\mathbf{b}$ its projection onto  a vector $\mathbf{a}$. 

Each image (state) $\mathcal{I}$ is characterized by the energy, $e_\mathcal{I}$, that for the regularized micromagnetic Hamiltonian is calculated using Eqs. (1) and (5) in the main text. 
After calculating the energies, one can determine the \textit{tangent} vectors:
\begin{eqnarray}
    \mathbf{T}_{\mathcal{I},i}^{*} = \begin{cases}
    \bm{\nu}_{\mathcal{I}+1,i}-\bm{\nu}_{\mathcal{I},i},\, e_{\mathcal{I}-1}<e_{\mathcal{I}}<e_{\mathcal{I}+1},\\
    \bm{\nu}_{\mathcal{I},i}-\bm{\nu}_{\mathcal{I}-1,i},\, e_{\mathcal{I}-1}>e_{\mathcal{I}}>e_{\mathcal{I}+1},\\
    \bm{\nu}_{\mathcal{I}+1,i}-\bm{\nu}_{\mathcal{I},i}+c(\bm{\nu}_{\mathcal{I},i}-\bm{\nu}_{\mathcal{I}-1,i}), \, e_{\mathcal{I}+1}>e_{\mathcal{I}-1},\\
    c(\bm{\nu}_{\mathcal{I}+1,i}-\bm{\nu}_{\mathcal{I},i})+\bm{\nu}_{\mathcal{I},i}-\bm{\nu}_{\mathcal{I}-1,i}, \, e_{\mathcal{I}+1}<e_{\mathcal{I}-1},
    \end{cases}\,\,\, c=\dfrac{\min\left(|e_{\mathcal{I}+1}-e_{\mathcal{I}}|,|e_{\mathcal{I}-1}-e_{\mathcal{I}}|\right)}{\max\left(|e_{\mathcal{I}+1}-e_{\mathcal{I}}|,|e_{\mathcal{I}-1}-e_{\mathcal{I}}|\right)},\label{tangent_vec1}
\end{eqnarray}
which must then be projected and normalized:
\begin{eqnarray}
\mathbf{T}_{\mathcal{I},i}=\mathcal{P}\left(\bm{\nu}_{\mathcal{I},i},\mathbf{T}_{\mathcal{I},i}^{*}\right),\,\,\,\bm{\tau}_{\mathcal{I},i}=\mathbf{T}_{\mathcal{I},i}\Big/\sqrt{\sum_{i=1}^{N}\left(\mathbf{T}_{\mathcal{I},i}\right)^{2}}.\label{tangent_vec2}
\end{eqnarray}
The spring force, which ensures the equidistant distribution of images along the path, is given by:
\begin{equation}
    \bm{f}_{\mathcal{I},i}=k(\rho_{\mathcal{I}+1,\mathcal{I}} - \rho_{\mathcal{I},\mathcal{I}-1})\ \bm{\tau}_{\mathcal{I},i},
\end{equation}
where $k$ is the spring stiffness constant.
The total force acting on the image $\mathcal{I}$ also includes components arising from the Hamiltonian, which have to be projected out into the tangent vectors:
\begin{equation}
\bm{g}_{\mathcal{I},i}=\mathcal{P}\left(\bm{\tau}_{\mathcal{I},i},\bm{\beta}_{\mathcal{I},i}\right),
\end{equation}
where $\bm{\beta}_{\mathcal{I},i}=-\dfrac{1}{M_\mathrm{s}}\dfrac{\delta \mathcal{H}}{\delta\bm{\nu}_{\mathcal{I},i}}$ is the effective field vector.
Thus, the total force in the RGNEB method is written as
\begin{equation}
    \mathcal{F}_{\mathcal{I},i} = \mathcal{P}\left(\bm{\nu}_{\mathcal{I},i},\bm{f}_{\mathcal{I},i}+\bm{g}_{\mathcal{I},i}\right).\label{total_force}
\end{equation}
In the case of endpoints ($\mathcal{I}=1$ and $\mathcal{I}=\mathcal{N}$), the force \eqref{total_force} includes the effective field only:
\begin{equation}
    \mathcal{F}_{\mathcal{I},i} = \mathcal{P}\left(\bm{\nu}_{\mathcal{I},i},\bm{\beta}_{\mathcal{I},i}\right).\label{total_force_endpoint}
\end{equation}
In both formulas, Eqs.~\eqref{total_force},~\eqref{total_force_endpoint}, the additional projection is used due to the convenience, as at the equilibrium points one has $\bm{\nu}_{\mathcal{I},i}\parallel\bm{\beta}_{\mathcal{I},i}$ or equivalently a zero force, $|\mathcal{F}_{\mathcal{I},i}|=0$.

The primary goal of the RGNEB is to minimize all forces, achieved using the algorithm and convergence criteria described in the next section.

\section{Velocity projection optimization}\label{AppendixC}
Denoting the set of all forces $\mathcal{F}=\{\mathcal{F}_\mathcal{I}, \mathcal{I}\in[1,\mathcal{N}]\}$ where  $\mathcal{F}_\mathcal{I}=\{\mathcal{F}_{i,\mathcal{I}}, i\in[1, N]\}$, the velocity projection optimization (VPO) represents the descent method, $\mathcal{M}\mapsto\mathcal{M}+\omega\mathcal{F}$, with an adaptive change of the weight parameter, $\omega$.
The convergence criterion for the method is defined simply as:
\begin{equation}
    |\mathcal{F}|=\dfrac{1}{N(\mathcal{N}-2)}\sqrt{\sum_{\mathcal{I}=2}^{\mathcal{N}-1}\sum_{i=1}^{N}\left(\mathcal{F}_{\mathcal{I},i}\right)^{2}}<\epsilon,
\end{equation}
meaning that the average force is less than a given small $\epsilon$.
The VPO algorithm is given as follows:
\begin{lstlisting}
INPUT:  magnetization array M0;  tolerance tol; step dt; mass m; 
        maximal number of steps MaxIter;
        
OUTPUT: magnetization array M1;

Step 1: Set velocity array to zero: V=0
Step 2: for i = 1,2,...,MaxIter do Step 3.
Step 3: Calculate forces, F, according  to Eq.(22) and Set ff = Dot(F,F)
        IF sqrt(ff)/size(M0) < tol  
            OUTPUT M0 and STOP
        ELSE do Steps 4-8
Step 4: Set vf = Dot(V+F*dt/2m,F)
	      IF vf <= 0 
            Set V = 0
	      ELSE 
            Set V = F*vf/ff
Step 5: Obtain the descent direction: D = V + F*dt/2m
Step 6: Perform the descent step:  M1 = M0 + dt*D
Step 7: Set velocity projected on M1:  V = D*Dot(M0,M1)-M0*Dot(M1,D)
Step 8: Set magnetization: M0 = M1 and  do Step 3
Step 9: PRINT(ERROR: The maximal number of iterations was exceeded!)
	      OUTPUT M0
Step 10:STOP.
\end{lstlisting}

In the above listing, we optimize the movable images only because the endpoint images are the local minimum states or have to be separately optimized according to any available minimization procedure.
To minimize the endpoints with the VPO method in parallel to the RGNEB calculations, one has to account for possible different values for the weight parameter $\omega$ for both subroutines.

\end{document}